\documentclass[preprint,aps,epsfig]{revtex4}
\usepackage{graphics,psfig,epsfig,graphicx}
\usepackage{amsbsy}
\usepackage{bm}
\usepackage{amsfonts}
\usepackage{amsmath}
\usepackage{cancel}
\usepackage{multirow}
\newcommand {\be}{\begin{equation}}
\newcommand {\ee}{\end{equation}}
\newcommand {\bea}{\begin{eqnarray}}
\newcommand {\ea}{\end{eqnarray*}}
\newcommand {\ba}{\begin{eqnarray*}}
\newcommand {\eea}{\end{eqnarray}}

\newcommand {\bra}{\langle}
\newcommand {\ket}{\rangle}

\newcommand {\refeq}[1] {(\ref{#1})}

\begin{document}
\title{Variational description of continuum states in terms of integral relations}

\author{A. Kievsky and M. Viviani}
\affiliation{Istituto Nazionale di Fisica Nucleare, Largo Pontecorvo 3, 56100 Pisa, Italy}
\author{Paolo Barletta}
\affiliation{Department of Physics and Astronomy, University College London,
Gower Street, London WC1E 6BT, United Kingdom}
\author{C. Romero-Redondo and E. Garrido}
\affiliation{Instituto de Estructura de la Materia, CSIC, Serrano 123, E-28006 
Madrid, Spain}

\begin{abstract}
Two integral relations derived from the Kohn Variational Principle (KVP)
are used for describing scattering states. 
In usual applications the KVP requires the
explicit form of the asymptotic behavior of the scattering wave function.
This is not the case when the integral relations are applied since,
due to their short range nature, the only condition for the scattering wave 
function $\Psi$ is that it be the solution of $(H-E)\Psi=0$ in the internal region. 
Several examples are analyzed for the computation of phase-shifts from bound state
type wave functions or, in the case of the scattering of charged particles,
it is possible to obtain phase-shifts using free asymptotic conditions. As a
final example we discuss the use of the integral relations in the case
of the Hyperspherical Adiabatic method.
\end{abstract}

\maketitle

\section{Introduction}

The study of bound and scattering states in few-nucleon systems gives valuable 
information regarding the underlying nuclear interaction. 
The fact that the spectrum of each hydrogen and helium isotope 
has only one bound state in the mass region $A=2-4$,
limits the applicability of bound state methods
to a few states in the study of these nuclei. In Ref.~\cite{nogga03} 
a detailed study
of the three-nucleon bound states has been done, whereas a similar analysis
in the case of $^4$He can be found in Refs.~\cite{nogga02,viviani05}. 
In recent years much of the study of the three-nucleon system has been done
in the three-nucleon continuum (see Refs.~\cite{gloeckle,kievsky01} and references therein).
Results in the four-body system have been obtained so far in the energy region 
below the three particle breakup~\cite{fisher06,arnas07}. 

Well established methods for treating both, bound and scattering states, 
regard the solution
of the Faddeev equations ($A=3$) or Faddeev-Yakubovsky equations ($A=4$)
in configuration or momentum space and the Hyperspherical Harmonic (HH)
expansion in conjunction with the Kohn Variational Principle (KVP). These
methods have proven to be of great accuracy. They have been tested using
different benchmarks~\cite{benchmark1,benchmark2}. 
On the other hand, many other methods
are presently used to describe bound states: for example the Green
Function Montecarlo (GFMC) and No Core Shell Model (NCSM) methods
have been used in nuclei up to $A=10$ and $A=12$ respectively~\cite{gfmc,ncsm}. 
Attempts to use these methods for the description of scattering
states have recently appeared~\cite{gfmc5,ncsm5}.

The possibility of employing bound state techniques for describing scattering states
has always attracted particular attention~\cite{harris}. Recently 
continuum-discretized states obtained from the stochastic variational method
have also been used to study $\alpha+n$ scattering~\cite{suzuki}. In those two
approaches the tangent of the phase-shift results to be a quotient of two numbers.
In the former the numerator and denominator are obtained from two
integral relations after projecting the Schr\"odinger equation,
whereas in the latter
the numerator results from an integral relation derived by means of the
Green's function formalism and the denominator from the
normalization of the continuum-discretized state. 

Another
problem that has received particular attention in few-nucleon scattering
processes regards collisions between charged particles. Traditionally,
the Faddeev method has been applied to the neutral $n-d$ reaction.
Applications to $p-d$ zero energy scattering were studied in configuration space
by the Los Alamos-Iowa group using $s$-wave potentials~\cite{friar83} and
realistic forces~\cite{friar84}. In those calculations the KVP was used to
correct the first order estimate of the scattering length after solving
the Faddeev equations in which the partial wave expansion of the Coulomb
potential was truncated. Low energy $p-d$ elastic scattering has also been
studied using the pair correlated hyperspherical harmonic (PHH) 
expansion~\cite{phh}.
A benchmark comparing these two techniques was given in Ref.~\cite{kievsky01b}. 
A different way to treat the Coulomb potential in few-nucleon scattering was 
proposed in Ref.~\cite{deltuva1}, based on the works of Ref.~\cite{alt},
in which the Alt-Grassberger-Sandhas equations were solved using
a screened Coulomb potential and then the scattering amplitude was obtained
after a renormalization procedure. Summarizing, the description of scattering states 
using very accurate methods are at present limited to $A\le4$ systems.
On the other hand, accurate methods for describing bound states beyond the $A=4$ mass
system exist, therefore the discussion of new methods for extending these approaches
to treating scattering states is of interest. In this discussion the treatment
of the Coulomb interaction cannot be neglected.
 
Recently two integral relations have been derived from the KVP~\cite{intrel}.
It has been shown that starting from the KVP, the tangent of the phase-shift
can be expressed as a quotient where the numerator and the denominator
are given in term of two integral relations. This is similar to what was proposed
in Ref.~\cite{harris}, however the variational character of the quotient and
its strict relation to the KVP were not recognized. 
In fact, it is this property that
makes possible many different and interesting applications of the integral relations.
Accordingly, in the present study we would like to discuss some specific examples.
We will show that the integral relations can be used to compute phase-shifts from
bound state like functions. 
We start our analysis from the simplest case, the $A=2$ system, using a model 
potential. Then, using a semirealistic
interaction, $n-d$ as well as $p-d$ scattering are considered. This is
of particular interest since, as we mentioned before, $p-d$ scattering has been
a subject of intense investigations. A second application of the integral
relations regards the possibility of determining
$p-d$ phase-shifts from a calculation in which the Coulomb 
potential has been screened. Finally, as a third application, we will discuss
the use of the integral relations with scattering wave functions obtained
from the Hyperspherical Adiabatic (HA) expansion. All these examples serve
to demonstrate the general validity of the KVP formulated in terms
of integral relations. Due to their short-range nature, they are determined
by the wave function in the interaction region and not from its
explicit asymptotic behavior. This means that any wave function $\Psi$
satisfying $(H-E)\Psi=0$ in the interaction region can be used to determine
the corresponding scattering amplitude even when its asymptotic behavior is not
the physical one.

The discussion presented here is limited to systems with $A=2,3$. This is because
our expertise to calculate few-nucleon wave functions is limited to these
systems. However applications to heavier systems are possible and, in particular,
it would be interesting to
analyze the use of the GFMC method in the computation of the integral relations
in systems with $A>4$. The paper is organized as follows: in Section II the integral
relations are derived from the KVP. Applications to the two- and three-body systems
are given in Section III and IV, respectively, whereas applications of the integral
relations in connection with the HA are given in Section V. The conclusions are given
in the last section.

\section{Integral Relations from the Kohn Variational Principle}
In order to derive the integral relations we
first consider a two-body system interacting 
through a short-range potential $V(r)$ at the center of mass energy $E$
in a relative angular momentum state $l=0$. 
The solution of the Schr\"odinger equation in configuration space
($m$ is twice the reduced mass),
\begin{equation}
(H-E)\Psi(\bm r)=(-\frac{\hbar^2}{m}\nabla^2+V-E)\Psi(\bm r)=0 \;\; ,
\label{eq1}
\end{equation}
can be obtained after specifying the corresponding boundary conditions.
For $E>0$, with $k^2=E/(\hbar^2/m)$ and assuming a short-range potential $V$, 
$\Psi(\bm r)=\phi(r)/\sqrt{4\pi}$ and
\begin{equation}
\phi(r\rightarrow\infty)\longrightarrow 
\sqrt{k} \left[A\frac{\sin(kr)}{kr}+B\frac{\cos(kr)}{kr}\right ] \;\; .
\end{equation}
from which one gets $\Psi \rightarrow A F +B G$, where
\begin{eqnarray}
 F&=&\sqrt{\frac{k}{4\pi}} \frac{\sin(kr)}{kr}  \nonumber \\
 G&=&\sqrt{\frac{k}{4\pi}} \frac{\cos(kr)}{kr} \;\; .
\end{eqnarray}
Use of the Wronskian theorem immediately leads to the following general expressions
for the coefficients $A$ and $B$:
\begin{eqnarray}
B& = & \frac{m}{\hbar^2}\left[<F|H-E|\Psi>- <\Psi|H-E|F>\right]  \nonumber \\
A& = & \frac{m}{\hbar^2}\left[<\Psi|H-E|G>- <G|H-E|\Psi>\right] \; \;,
\label{wrons}
\end{eqnarray}
where we have made use of the fact that:
\begin{equation}
\frac{m}{\hbar^2}\left[<F|H-E|G>-<G|H-E|F>\right] =1 \;\; .
\label{norm0}
\end{equation}
With the above normalization, and assuming that $\Psi$ is an exact solution of
Eq.(\ref{eq1}), it follows that $\Psi$ satisfies the following
integral relations:
\begin{equation}
\begin{aligned}
-&\frac{m}{\hbar^2} <\Psi|H-E|F>=B  \cr
&\frac{m}{\hbar^2} <\Psi|H-E|G>=A  \cr
& \tan\delta = \frac{B}{A} \;\; .
\end{aligned}
\label{rel1}
\end{equation}
Explicitly they are
\begin{equation}
\begin{aligned}
-& \frac{m}{\hbar^2\sqrt{k}}\int_0^\infty dr 
\sin(kr)V(r)[r\phi(r)]=B  \cr
 & \frac{m}{\hbar^2\sqrt{k}}\int_0^\infty dr 
\cos(kr)V(r)[r\phi(r)]+\frac{\phi(0)}{\sqrt{k}} =A  \;\; ,
\end{aligned}
\label{origin}
\end{equation}
where in the last integral we have used the property that
$\nabla^2(1/r)=-4\pi\delta({\bm r})$.

In practical cases the solution of the Schr\"odinger equation is obtained
numerically. Then, $\tan\delta$ is extracted from $\phi(r)$ analyzing its 
behavior
outside the range of the potential. The equivalence between the extracted
value and that obtained from the integral relations defines the accuracy
of the numerical computation. A relative difference of the order of
$10^{-7}$ of the two values is usually achieved using standard numerical
techniques to solve the differential equation and to compute the two 
one-dimensional integrals. The short range character of the
integral relations should be noticed. 
This means that the phase-shift is determined by
the internal structure of the wave function.

The second relation of Eq.~\refeq{origin} shows a dependence on the value of the 
wave function at the origin. It could be convenient to eliminate this explicit
dependence since the numerical determination of $\phi(0)$ might be problematic,
as we will show. To this end we introduce a regularized function
$\tilde G=f_{reg}G$ with the property that $|\tilde G(r=0)|<\infty$ and
$\tilde G=G$ outside the interaction region. A possible choice is
\begin{equation}
 \tilde G=\sqrt{\frac{k}{4\pi}}\frac{\cos(kr)}{kr}(1-{\rm e}^{-\gamma r})\;\; ,
\end{equation}
where the regularization function $f_{reg}=(1-{\rm e}^{-\gamma r})$ has been 
introduced where
$\gamma$ is a non linear parameter which will be discussed below. 
Values satisfying $\gamma>1/r_0$,
with $r_0$ the range of the potential, could be appropriate.
The regularized function $\tilde G$ (as well as the irregular function $G$), 
satisfies the normalization condition
\begin{equation}
\frac{m}{\hbar^2}\left[<F|H-E|\tilde G>-<\tilde G|H-E|F>\right] =1 \;\; .
\label{norm}
\end{equation}
Therefore the second integral relation in Eq~\refeq{rel1} remains valid using
$\tilde G$ in place of $G$,
\begin{equation}
 \frac{m}{\hbar^2} <\Psi|H-E|\tilde G>=A \;\; ,
\label{reg1}
\end{equation}
with the explicit form:
\begin{equation}
  \frac{m}{\hbar^2\sqrt{k}}\int_0^\infty d{r} 
 \cos(kr)V(r)[r\phi(r)]+ I_\gamma =A  
\label{reg2}
\end{equation}
where in $I_\gamma$ all terms depending on $\gamma$, introduced by 
$f_{reg}$, are included:
\begin{equation}
 I_\gamma= -\frac{1}{\sqrt{k}}\int_0^\infty dr 
 \left(\frac{m}{\hbar^2}V(r)\cos{kr}-\gamma^2\cos{kr}-2\gamma k\sin{kr}
 \right){\rm e}^{-\gamma r}[r\phi(r)]
\end{equation}
Comparing Eq.~\refeq{reg2} to Eq.~\refeq{origin} we identify $I_\gamma=\phi(0)/\sqrt{k}$.
This equality can be verified with the same
relative accuracy obtained for $\tan\delta$ provided that the regularization
of $G$ is done inside the interaction region. 

In the following we demonstrate that the relation $\tan\delta=B/A$, which is
an exact relation when the exact wave function $\Psi$ is used in Eq~\refeq{rel1}, 
can be considered accurate up to second order when a trial wave function is
used, as it has a strict connection with the Kohn variational principle. 

The connection of the integral relations with the KVP is straightforward.
Defining a trial wave function $\Psi_t$  to be
\begin{equation}
 \Psi_t=\Psi_c +AF+B\;\tilde G  \;\; ,
\label{psic}
\end{equation}
with $\Psi_c\rightarrow 0$ as
$r\rightarrow\infty$, the condition
$\Psi_t\rightarrow A F+B \; G$ as $r\rightarrow\infty$ is fulfilled. The KVP
states that the second order estimate for $\tan\delta$ is
\begin{equation}
[\tan\delta]^{2^{nd}}=\tan\delta - \frac{m}{\hbar^2}<(1/A)\Psi_t|H-E|(1/A)\Psi_t> \,\ .
\label{kohn}
\end{equation}
The above functional is stationary with respect to variations
of $\Psi_c$ and $\tan\delta$. Without a loss of generality $\Psi_c$ can be
expanded in terms of a (square integrable) complete basis
\begin{equation}
 \Psi_c=\sum_n a_n \phi_n(r) \;\; .
\end{equation} 
The variation of the functional with respect to the linear
parameters $a_n$ 
and $\tan\delta$ leads to the following equations
\begin{equation}
\begin{aligned}
 &<\phi_n|H-E|\Psi_t>=0  \cr
 &<\tilde G|H-E|\Psi_t>=0  \;\;\; .
\end{aligned}
\label{first}
\end{equation}
To obtain the last equation, the normalization relation of Eq.~\refeq{norm} 
has been used.
From these two equations, $\Psi_c$ and the first order
estimate of the phase shift $(\tan\delta)^{1^{st}}$ can be determined. 
It should be noted that the first equation implies 
$<\Psi_c|H-E|\Psi_t>=0$. Furthermore, from the general relation for
$A$ in Eq.~\refeq{wrons}, and using 
the second equation in Eq.~\refeq{first}, the following integral relation results
\begin{equation}
\frac{m}{\hbar^2}<\Psi_t|H-E|\tilde G>=A  \;\; .
\end{equation}

Replacing the two relations of Eq.\refeq{first} into the functional of 
Eq.\refeq{kohn}, a second order estimate of the phase shift is obtained
\begin{equation}
[\tan\delta]^{2^{nd}}=(\tan\delta)^{1^{st}} 
- \frac{m}{\hbar^2}<F|H-E|(1/A)\Psi_t> \,\ .
\label{second1}
\end{equation}
Multiplying Eq.~\refeq{second1} by $A$ one gets
\begin{equation}
B^{2^{nd}}=B^{1^{st}} 
- \frac{m}{\hbar^2}<F|H-E|\Psi_t> \,\ .
\label{second2}
\end{equation}
On the other hand, a first order estimate for the coefficient $B$ can be obtained
from the general relation in Eq.~\refeq{wrons}, i.e.,
\begin{equation}
\frac{m}{\hbar^2}\left[<F|H-E|\Psi_t>-
           <\Psi_t|H-E|F>\right]=B^{1^{st}} \,\,\ .
\label{firstb}
\end{equation}
Therefore,
replacing Eq.\refeq{firstb} in Eq.\refeq{second2}, a second order
integral relation for $B$ is obtained. The above results can be
summarized as follow
\begin{equation}
\begin{aligned}
 B^{2^{nd}}& = & -\frac{m}{\hbar^2}<&\Psi_t|H-E|F>   \cr
 A & = & \frac{m}{\hbar^2}<&\Psi_t|H-E|\tilde G>  \cr
 [\tan\delta]^{2^{nd}} & = & & B^{2^{nd}}/A  \,\, .
\end{aligned}
\label{relint}
\end{equation}

These equations extend the validity of the integral relations, 
given in Eq.\refeq{rel1} for the exact wave functions, to trial
wave functions. To be noticed that $F,\tilde G$ are solutions
of the Schr\"odinger equation in the asymptotic region, therefore
$(H-E)F\rightarrow 0$ and $(H-E)\tilde G\rightarrow 0$ 
as the distance between the particles increases. 
As a consequence the decomposition of $\Psi_t$ in
the three terms of Eq.~\refeq{psic} can be considered formal since,
due to the short-range character
of the integral relations, it is sufficient for the trial wave function
to be a solution of $(H-E)\Psi_t=0$ in the interaction region, without
an explicit indication of its asymptotic behavior. This fact,
together with the variational character
of the relations, allows for a number of applications to be discussed
in the next sections.

\section{Use of the Integral Relations in the two-body case}

In this section we present applications of the integral relations of
Eq.~\refeq{relint} to a two-body system. To make contact with the
results given in Refs.~\cite{intrel,barletta09}, we use a central, 
$s$-wave gaussian potential
\begin{equation}
 V(r)=-V_0\exp{(-r^2/r_0^2)}  \;\; ,
\end{equation}
with $V_0=-51.5$ MeV, $r_0=1.6$ fm and $\hbar^2/m=41.4696$ MeV fm$^2$.
This potential has a shallow $L=0$ bound state with energy 
$E_{2B}=-0.397743$ MeV.

We introduce the orthogonal basis
\begin{equation}
 \phi_m={\cal L}_m^{(2)}(z)\exp{-(z/2)}  \;\; ,
\end{equation}
with ${\cal L}_m$ a (normalized) 
Laguerre polynomial and $z=\beta r$, where $\beta$ is
a nonlinear parameter, to expand the wave function of the system
\begin{equation}
 \Psi_0=\sum_{m=0}^{M-1} a^0_m \phi_m  \,\, .
\end{equation}
We solve the eigenvalue problem of $H$ for different dimensions $M$ of the 
basis. The variational principle states that
\begin{equation}
E_0=\bra \Psi_0|H|\Psi_0 \ket \ge E_{2B}  \;\; ,
\end{equation}
with the equality valid when $M\rightarrow\infty$. 
The nonlinear parameter $\beta$ can be fixed to improve the
convergence properties of the basis. In fact, 
for each value of $M$ there is a value of $\beta$ that minimizes the energy.
Increasing $M$, the minimum of the energy becomes less and less dependent
on $\beta$ resulting in a plateau.
Increasing further the dimension of the basis, the extension
of the plateau increases as well, without any appreciable improvement in
the eigenvalue, indicating that the convergence has been reached to a
certain accuracy. At each step $\Psi_0$
represents a first order estimate of the exact bound state wave function. 

\begin{figure}
\vspace{1.0cm}
\epsfig{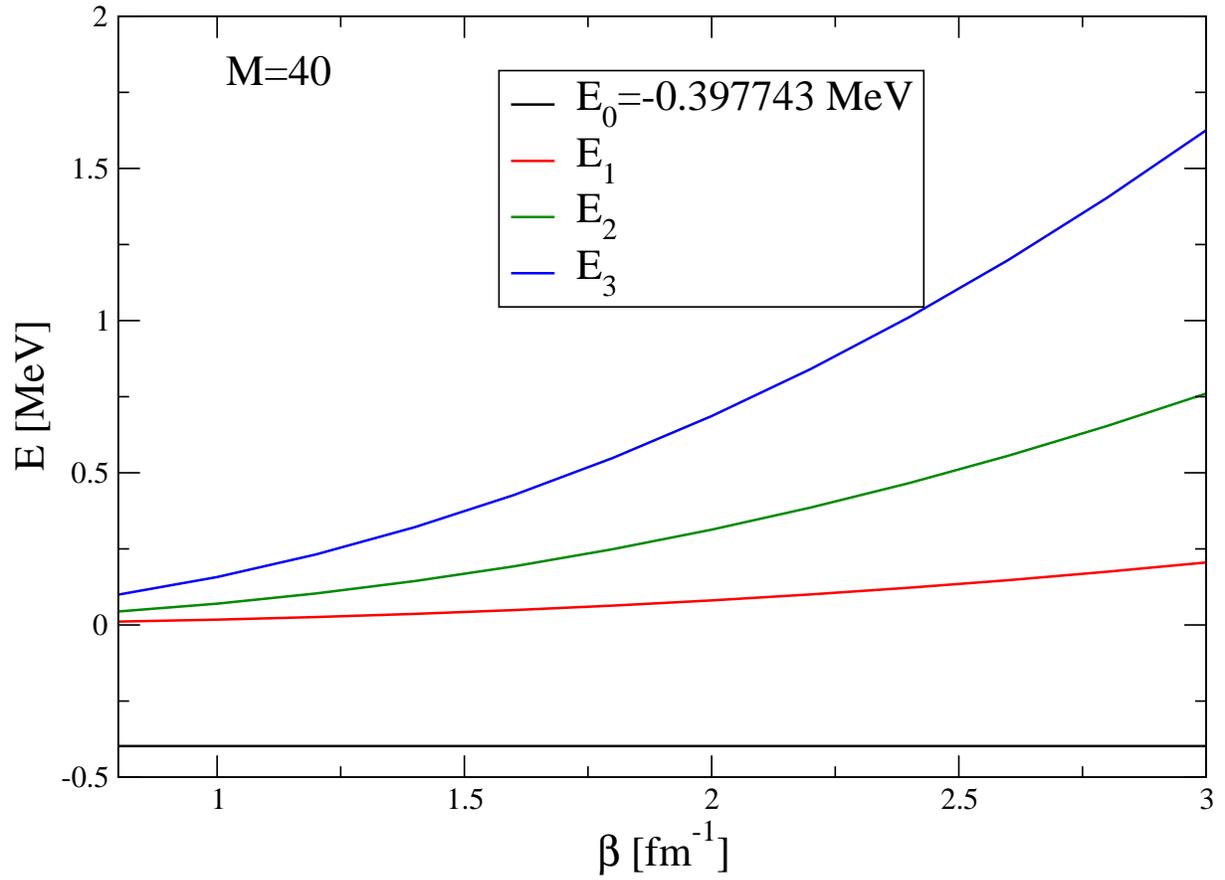}
\caption{(Color on line) 
The two-nucleon bound state energy $E_0$ and the first three positive 
 eigenvalues $E_j$ as a function of $\beta$ in the case of $M=40$}
\label{fig:fig1}
\end{figure}

Since, in our
example, the system has only one bound state, with appropriate values of $M$ and $\beta$,
the diagonalization of $H$ results in one negative eigenvalue
$E_0$ and $M-1$ positive eigenvalues $E_j$ ($j=1,....,M-1$). The
corresponding wave functions 
\begin{equation}
 \Psi_j=\sum_{m=0}^{M-1} a^j_m \phi_m  \hspace{0.5cm} j=1,....,M-1 \;\; ,
\end{equation}
are approximate solutions of $(H-E_j)\Psi_j=0$ in the interaction region. 
As $r\rightarrow\infty$ 
they go to zero exponentially and therefore they do not represent physical 
scattering states.
The negative energy $E_0$ and the first three positive energy eigenvalues
($E_j$, $j=1,3$)
are shown in Fig.~\ref{fig:fig1} as a function of $\beta$ for $M=40$.
We observe the plateau already reached by $E_0$ for the values
of $\beta$ showed in the figure. Furthermore, we observe the monotonic
behavior of the positive eigenvalues towards zero as $\beta$ decreases.
The corresponding eigenvectors
can be used to compute the integral relations of Eq.~\refeq{relint} and to 
calculate the second order estimate of the phase-shifts 
$\delta_j$ at the specific energies $E_j$. 
This analysis is shown in Table~\ref{tab:tab1} in which the non linear 
parameter $\beta$ of the Laguerre basis has been chosen to be $1.2$ fm$^{-1}$. 
In the first row of the
table the ground state energy is given for different values of the
number $M$ of Laguerre polynomials. The stability of $E_0$ at the
level of $1$ keV is achieved already with $M=20$. 
For a given value of $M$, $E_j$, with $j=1,2,3$, 
are the first three positive eigenvalues. 
The eigenvectors corresponding to positive energies approximate
the scattering states at these specific energies. Since the lowest
scattering state appears at zero energy, none of the positive eigenvalues
can reach this value for any finite values of $M$. We observe (see Fig.~\ref{fig:fig1}) 
that the eigenvalues diminish as $M$ increases.
Defining
$ k^2_j=\frac{m}{\hbar^2}E_j$, the second order estimate for the phase
shift at each energy and at each value of $M$ is obtained as
\begin{equation}
\begin{aligned}
-&\frac{m}{\hbar^2} <\Psi_j|H-E|F_j>=B_j
 &{\rm with} \hspace{1cm} F_j=\sqrt{\frac{k_j}{4\pi}} \frac{\sin(k_jr)}{k_jr}  \cr
&\frac{m}{\hbar^2} <\Psi_j|H-E|\tilde G_j>=A_j           
 &{\rm with} \hspace{1cm} \tilde G_j=f_{reg}\sqrt{\frac{k_j}{4\pi}} 
\frac{\cos(k_jr)}{k_jr}  \cr
& [\tan\delta_j]^{2^{nd}} = \frac{B_j}{A_j}.
\end{aligned}
\label{rel2}
\end{equation} 

On the other hand, as we are considering the $A=2$ system,
at each specific energy value $E_j$ the phase shift $\tan\delta_j$ 
can be obtained
by solving the Schr\"odinger equation numerically. The two values,
$[\tan\delta_j]^{2^{nd}}$ and $\tan\delta_j$, are given in the Table~\ref{tab:tab1}
 at the corresponding energies as a function of $M$. We observe that, as
$M$ increases, the relative difference between the variational estimate 
and the exact value reduces, for example at $M=40$ it is about $10^{-6}$. In fact,
as $M$ increases, each eigenvector gives a better representation of
the exact wave function in the internal region and the second order
estimates, $[\tan\delta_j]^{2^{nd}}$ approach the exact result. 

\begin{figure}
\epsfig{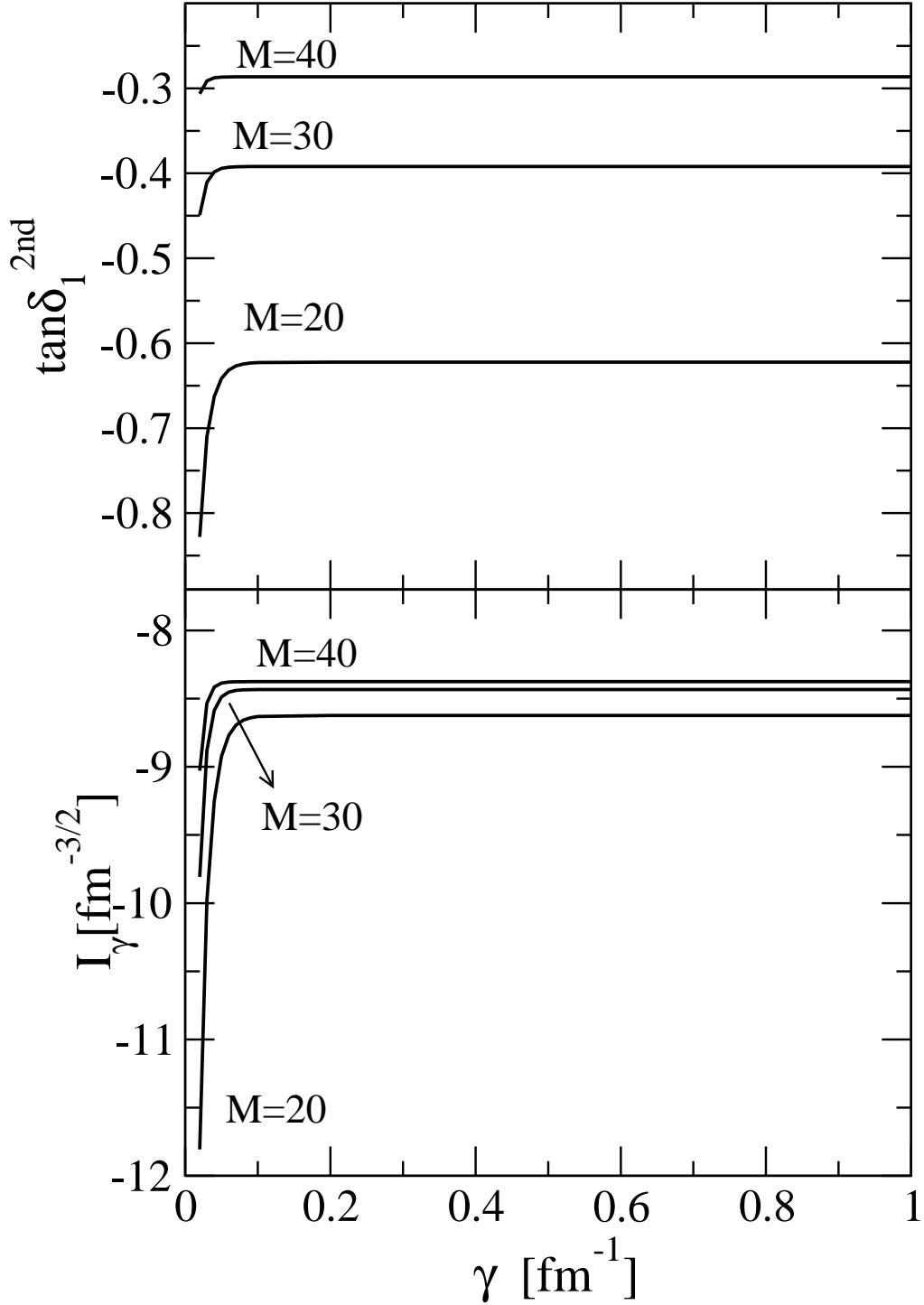}
\caption{The two-nucleon second order estimate, 
\protect$[\tan\delta_1]^{2^{nd}}$, calculated
using $\Psi_1$ as a function of the non linear parameter $\gamma$,
for the values $M=20,30,40$ and the integral $I_\gamma$ as a function
of $\gamma$ at the same three values of $M$.}
\label{fig:fig2}
\end{figure}

The study of the stability of the results in terms of the non linear
parameter $\gamma$ in the regularization function $f_{reg}$
is given in Fig.~\ref{fig:fig2}. In the upper panel, the second order
$[\tan\delta_1]^{2^{nd}}$, corresponding to the eigenvalue $E_1$ given
in Table~\ref{tab:tab1}, is shown as a function of $\gamma$, for
$M=20,30,40$. We observe a good stability for values of 
$\gamma > 0.2\;{\rm fm}^{-1}$ indicating that the regularization has
to be done before $\approx 5\;$fm. In lower panel the corresponding
values for $I_\gamma$ as a function
of $\gamma$ are shown. The stable values obtained for $M=20,30,40$ 
when $\gamma > 0.2\;{\rm fm}^{-1}$ are $I_\gamma=-8.6234,-8.4334,-8.3755$
respectively. The exact values for $\phi(0)/\sqrt{k}$, obtained solving
the Schr\"odinger equation numerically at the three energies are:
$-8.6188,-8.4338,-8.3755$, respectively. We can observe that for
$M=40$ there is a complete agreement between $I_\gamma$ and
$\phi(0)/\sqrt{k}$. Therefore $I_\gamma$ can be considered to be an integral
representation of $\phi(0)/\sqrt{k}$. This is an important point since
such a value can be used to normalize the variational wave function.
In this example the integral relations derived from the KVP
have been used to compute phase-shifts using bound state like
wave functions.

\begin{figure}
\epsfig{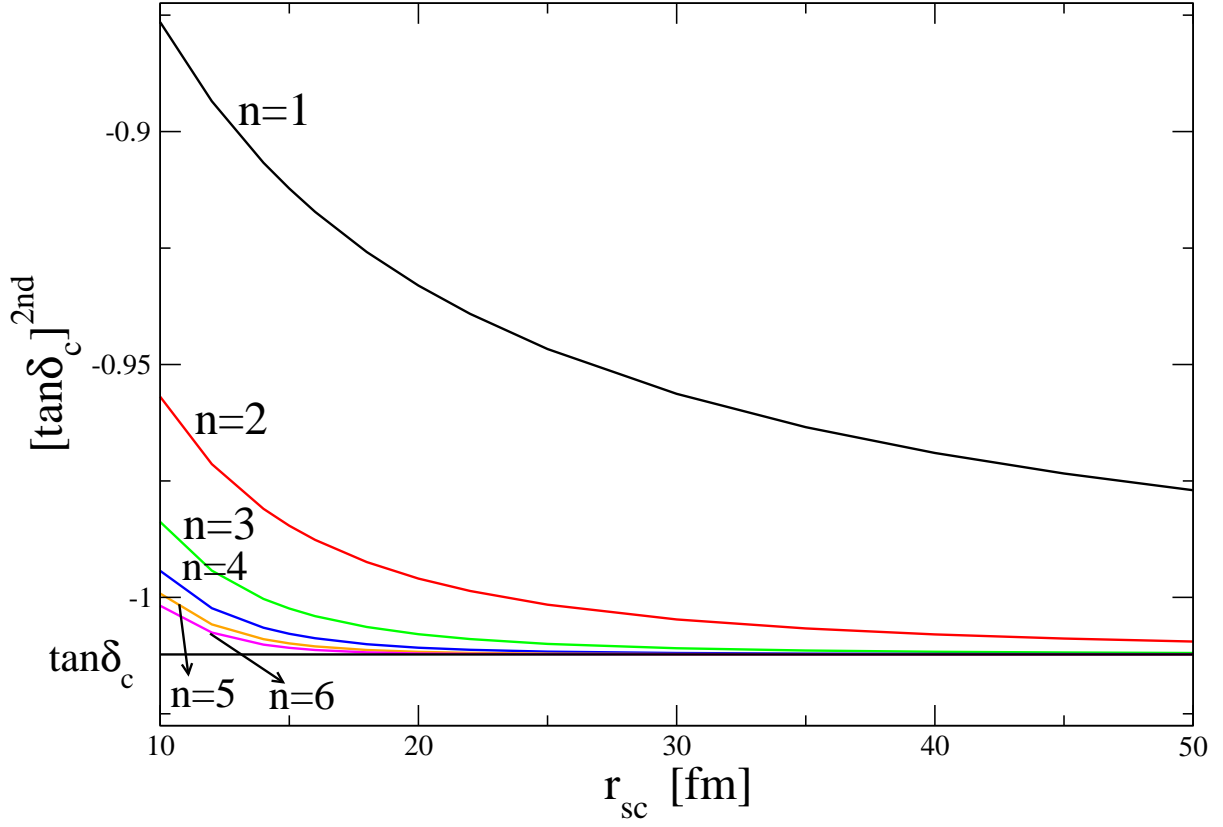}
\caption{(Color on line)
The two-nucleon second order estimate $[\tan\delta_c]^{2^{nd}}$
as a function of $r_{sc}$ for different values of $n$. As a reference
the exact value for $\tan\delta_c$ is given as a straight line.
}
\label{fig:fig3}
\end{figure}

A different application of the integral relations regards the possibility
of calculating the phase-shift of a process in which the two particles
interact through a short range potential plus the Coulomb potential, 
imposing free asymptotic conditions to the wave function. 
As an example we use the same two body potential used in the previous analysis
and add the Coulomb potential:
\begin{equation}
 V(r)=-V_0\exp{-(r/r_0)^2}+ \frac{e^2}{r}  \,\,\, .
\label{potc}
\end{equation}
For positive energies and $l=0$, the wave function behaves asymptotically as
\begin{equation}
 \Psi^{(c)}(r\rightarrow\infty)= AF_c(r)+BG_c(r)\;\; ,
\label{asympc}
\end{equation}
with $F_c(r),G_c(r)$ the regular and irregular Coulomb functions, respectively. 
The phase-shift is $\tan\delta_c=B/A$. The KVP remains valid when the long range
Coulomb potential is considered and its form in terms of the integral
relations results to be:
\begin{equation}
\begin{aligned}
-&\frac{m}{\hbar^2} <\Psi^{(c)}_t|H-E|F_c>=B  \cr
&\frac{m}{\hbar^2} <\Psi^{(c)}_t|H-E|\tilde G_c>=A            \cr
& [\tan\delta_c]^{2^{nd}} = \frac{B}{A}\;\; .
\end{aligned}
\label{rel3}
\end{equation}
with $\tilde G_c=f_{reg}G_c$ and $\Psi^{(c)}_t$ a trial wave function
behaving asymptotically as $\Psi^{(c)}$. Since $(H-E)|F_c>$ and $(H-E)|\tilde G_c>$ go to
zero outside the range of the short range potential, the integrals in
Eq.~\refeq{rel3} are negligible outside that region. Therefore, for
the computation of the phase-shift it is enough to require that $\Psi^{(c)}_t$ satisfies
$(H-E)\Psi^{(c)}_t=0$, inside that region. To exploit this fact, we introduce the 
following screened potential:
\begin{equation}
 V_{sc}(r)=-V_0\exp{[-(r/r_0)^2]}+ \left[{\rm e}^{-(r/r_{sc})^n}\right]\frac{e^2}{r}\;\; .
\end{equation}
For specific values of $n$ and $r_{sc}$ it has the property of being
extremely close to the potential $V(r)$ of Eq.~\refeq{potc} 
for $r<r_0$, with $r_0$ the range of the short
range potential. The screening factor ${\rm e}^{-(r/r_{sc})^n}$ cuts the 
Coulomb potential for $r>r_{sc}$. 
Using the potential $V_{sc}$ to describe a scattering process, 
the wave function behaves asymptotically as
\begin{equation}
 \Psi_{n,r_{sc}}(r\rightarrow\infty)= F(r)+\tan\delta_{n,r_{sc}}\;G(r)
\end{equation}
where $F,G$ are given by Eq.~\refeq{rel2},
since $V_{sc}$ is a short range potential. It should be noted that the screened
phase-shift $\tan\delta_{n,r_{sc}}$ does not equal $\tan\delta_c$ for any finite value
of $n$ and $r_{sc}$.
Solving the Schr\"odinger equation for $V_{sc}$,
it is possible to obtain the wave function $\Psi_{n,r_{sc}}$
for different values of $n$ and $r_{sc}$.
This wave function can be considered to be a trial wave function for the problem
in which the Coulomb potential is unscreened. Accordingly it can be used as input 
in Eq.~\refeq{rel3} to obtain a second order estimate
of the Coulomb phase-shift,
\begin{equation}
\begin{aligned}
-&\frac{m}{\hbar^2} <\Psi_{n,r_{sc}}|H-E|F_c>=B  \cr
&\frac{m}{\hbar^2} <\Psi_{n,r_{sc}}|H-E|\tilde G_c>=A            \cr
& [\tan\delta_c]^{2^{nd}} = \frac{B}{A}
\end{aligned}
\label{rel4}
\end{equation}
where the unscreened Coulomb potential is included in $H$.
This estimate depends
on $n$ and $r_{sc}$ as the wave function does. In Fig.~\ref{fig:fig3} the
second order estimate $[\tan\delta_c]^{2^{nd}}$ is shown as a function of
$r_{sc}$ for different values of $n$. The straight line is the exact value
of $\tan\delta_c$ obtained solving the Schr\"odinger equation. 
We can observe that for $n\ge4$ and $r_{sc}>30$ fm
the second order estimate coincides with the exact results. In this example
the integral relations derived from the Kohn Variational Principle have
been used to extract a phase-shift in the presence of the Coulomb potential
using wave functions with free asymptotic conditions.

\section{Use of the Integral Relations in the three-body case}

The integral relations derived
from the KVP are general and their validity is not limited to two-body systems.
The two-body system is the simplest system in which different applications 
can be studied and compared to the exact solution of the Schr\"odinger
equation and, therefore,
a detailed analysis of the variational character of the relations can be
performed. In this section the study of the integral relations is extended to 
describe a $2+1$ collision in the three-body system, below the breakup threshold
into three particles. The description of the breakup channel remains outside
the scope of the present work. In the following
we will consider the two examples already discussed in the
previous section:
the computation of phase-shifts using bound state like wave functions and
the calculation of phase-shifts in presence of the Coulomb potential using
wave functions having free asymptotic conditions. 
To this end we will use the $s$-wave MT I-III nucleon-nucleon 
interaction~\cite{mtiii},
active in the singlet and triplet spin states, respectively: 
\be
\begin{aligned}
V_{MT\;I}(r) = & 
\frac{1438.72}{r} {\rm e}^{-3.11 r} - \frac{513.968}{r} {\rm e}^{-1.55 r}  \cr
 & \cr
V_{MT\;III}(r) = & 
\frac{1438.72}{r} {\rm e}^{-3.11 r} - \frac{626.885}{r} {\rm e}^{-1.55 r}
\end{aligned}
\ee
with distances in fm and energies in MeV.
This interaction has been used many times in the 
literature to study the three-nucleon system at low energies. 
It is considered a semi-realistic
interaction since it describes reasonably well the deuteron binding energy and
the singlet and triplet $n-p$ scattering lengths. Its predictions for these 
quantities are
$E_d=-2.23069$ MeV, ${}^1a_{n-d}=-23.582$ fm, and ${}^3a_{n-d}=5.5132$ fm. 
To be noticed that this potential has a strong repulsion at short distances. 

To compute bound and scattering wave functions we make use of 
the pair hyperspherical harmonic (PHH) method which has proven
to be extremely accurate~\cite{phh,phh1}. 
In the following a brief illustration of the
method is given. For bound states, the three-nucleon wave function
is decomposed in three Faddeev-like amplitudes
\be
\Psi=\psi({\bm X}_i,{\bm Y}_i)+\psi({\bm X}_j,{\bm Y}_j)
 +\psi({\bm X}_k,{\bm Y}_k) \;\; ,
\label{psi3b}
\ee
where we have introduced the Jacobi coordinates:
${\bm X}_i =  ({\bm r}_j - {\bm r}_k)/\sqrt{2}$ and
${\bm Y}_i =  ({\bm r}_j + {\bm r}_k - 2 {\bm r}_i)/\sqrt{6}$ (the generic vector
${\bm r}_k$ indicates the position of nucleon $k$). Each amplitude having
quantum numbers $J,J_z,T,T_z$ is expanded in angular-spin-isospin
channels (called $\alpha$-channels) as
\be
\begin{aligned}
\psi({\bm X}_i,{\bm Y}_i)= & \sum_\alpha \Phi_\alpha(X_i,Y_i) {\cal Y}_\alpha(jk,i) \cr
{\cal Y}_\alpha(jk,i)= & \{[Y_{l_\alpha}({\hat X}_i)\otimes
Y_{L_\alpha}({\hat Y}_i)]_{\Lambda_\alpha}[s^{jk}_\alpha \otimes
\textstyle{\frac{1}{2}}]_{S_\alpha}\}_{JJ_z} 
[t^{jk}_\alpha \otimes\textstyle{\frac{1}{2}}]_{TT_z}
\end{aligned}
\ee
and the radial amplitudes are expanded in terms of the PHH basis
\be
\Phi_\alpha(X_i,Y_i)=\rho^{l_\alpha+L_\alpha}f_\alpha(\sqrt{2}X_i)
\sum_K u_K^\alpha(\rho)\; ^{(2)}P_K^{l_\alpha,L_\alpha}(\phi_i)
\ee
where we have introduced the hyperspherical variables, the hyperradius $\rho$
and the hyperangle $\phi_i$, defined by the relations
$X_i=\rho\cos\phi_i, Y_i=\rho\sin\phi_i$, and $^{(2)}P_n^{l_\alpha,L_\alpha}(\phi_i)$
is a hyperspherical polynomial. The summation is given in terms of
the grand angular quantum number $K=2n+l_\alpha+L_\alpha$.
The correlation functions $f_\alpha(r)$ are
introduced to accelerate the rate of convergence of the expansion. They take into
account those configurations in which two nucleons are close to each other. A very
convenient choice is to derive the correlation functions from a Schr\"odinger like
equation governed by the two-body potential corresponding to the specific $\alpha$-channel 
\cite{phh1}. 

In the following we consider a three-nucleon system in either the
$J=1/2^+$ or $J=3/2^+$ states with total isospin $T=1/2$. The central character of the
MT I-III interaction decouples those channels with different values of 
$\Lambda_\alpha$. Moreover, 
as the interaction acts only in the $s$-wave, we have $l_\alpha=0$. This 
condition limits
the number of channels of the ($\Lambda_\alpha=0$) $J=1/2^+$ state to two
channels, corresponding to $s^{jk}_\alpha=0,1$, and to one channel, 
corresponding to $s^{jk}_\alpha=1$, in the case of the
($\Lambda_\alpha=0$) $J=3/2^+$ state. Finally, following
Ref.~\cite{kiev97}, the hyperradial functions are expanded in terms of
Laguerre polynomials 
\be
    u^\alpha_K(\rho)=\sum_{m=0}^{M-1} 
A^\alpha_{K,m} {\cal L}^{(5)}_m(z) \exp{(-z/2)}
\label{lag3}
\ee
with $z=\beta\rho$, and $\beta$ a non linear parameter. We can define
a complete antisymmetric three-nucleon state $|\alpha,K,m>$, in terms of which
the wave function $\Psi_n$ for the $n$-th state results to be
\be
    \Psi_n=\sum_{\alpha,K,m} A^{\alpha,n}_{K,m} |\alpha,K,m> \;\; .
\ee
The linear coefficients in the expansion are determined by solving the 
generalized eigenvalue problem
\be
    \sum_{\alpha',K',m'} A^{\alpha',n}_{K',m'} <\alpha,K,m|H-E_n|\alpha',K',m'>=0 \;\; .
\label{geneig}
\ee

The extension of the PHH method to describe scattering states below the deuteron breakup,
using the KVP, is straightforward~\cite{phh,kiev97}. 
As for bound states, we limit the discussion to the 
($\Lambda_\alpha=0$) $J=1/2^+,3/2^+$ states with total isospin $T=1/2$. 
The $N-d$ scattering wave $\Psi_k$ function at the center of mass energy
$E=E_d+(4/3)(\hbar^2/m)k^2$, is written as
\be
\begin{aligned}
 \Psi_k= & \sum_{\alpha,K,m} A^{\alpha,k}_{K,m} |\alpha,K,m> + |F_k> +
\tan\delta \; |\tilde G_k> \cr
|F_k>= & \sum_i \phi_d(X_i) F_0(k y_i)
[s^{jk} \textstyle{\frac{1}{2}}]_{JJ_z} 
[t^{jk} \otimes\textstyle{\frac{1}{2}}]_{TT_z} \cr
|\tilde G_k>= & \sum_i \phi_d(X_i) f_{reg}(y_i)
G_0(k y_i)
[s^{jk} \textstyle{\frac{1}{2}}]_{JJ_z} 
[t^{jk}_\alpha \otimes\textstyle{\frac{1}{2}}]_{TT_z} \;\; ,
\end{aligned}
\label{asym1}
\ee
with $\phi_d(X_i)$ the deuteron wave function having spin $s^{jk}=1$ and
isospin $t^{jk}=0$. $F_0,G_0$ are proportional to the regular and irregular 
Bessel functions in the case of $n-d$ scattering or to the regular and irregular 
Coulomb functions, divided by $ky_i$, in the case of $p-d$ scattering. 
The distance between the nucleon $i$
and the deuteron, formed by nucleons $j,k$, is $y_i$ and
$f_{reg}(y)=(1-\exp{(-\gamma y)})$ is the chosen regularization factor. 
In our calculations value of
$\gamma=0.25\;$fm$^{-1}$ has been found to be appropriate. 
The coefficients $A^{\alpha,k}_{K,m}$ and the first order estimate
of $\tan\delta$ are obtained by solving the following linear
system
\begin{equation}
\begin{aligned}
 & \sum_{\alpha',K',m'} A^{\alpha',k}_{K',m'} 
 <\alpha,K,m|H-E|\alpha',K',m'>+ \tan\delta<\alpha,K,m|{\cal G}_k >=
 -<\alpha,K,m|{\cal F}_k>   \cr
 & \sum_{\alpha,K,m} A^{\alpha,k}_{K,m}<\alpha,K,m|{\cal G}_k>
+\tan\delta<\tilde G_k|{\cal G}_k > =-<\tilde G_k|{\cal F}_k> \;\; ,
\end{aligned}
\label{linears}
\end{equation}
where we have defined $|{\cal G}_k>=(H-E)|\tilde G_k>$ and
$|{\cal F}_k>=(H-E) |F_k>$. Following Eq.~\refeq{relint}, the second order 
estimate for $\tan\delta$ is
\begin{equation}
\begin{aligned}
 B^{2^{nd}}_k& = & -\frac{m}{\hbar^2}<&\Psi_k|{\cal F}_k>   \cr
 A_k & = & \frac{m}{\hbar^2}<&\Psi_k|{\cal G}_k>  \cr
 [\tan\delta_k]^{2^{nd}} & = & & B^{2^{nd}}_k/A_k  \,\, .
\end{aligned}
\label{relintp}
\end{equation}
It should be observed that in the present case, due to the definition of the asymptotic
behavior of $\Psi_k$, we have $(m/\hbar^2)<\Psi_k|{\cal G}_k>=1$.

\begin{figure}

\epsfig{file=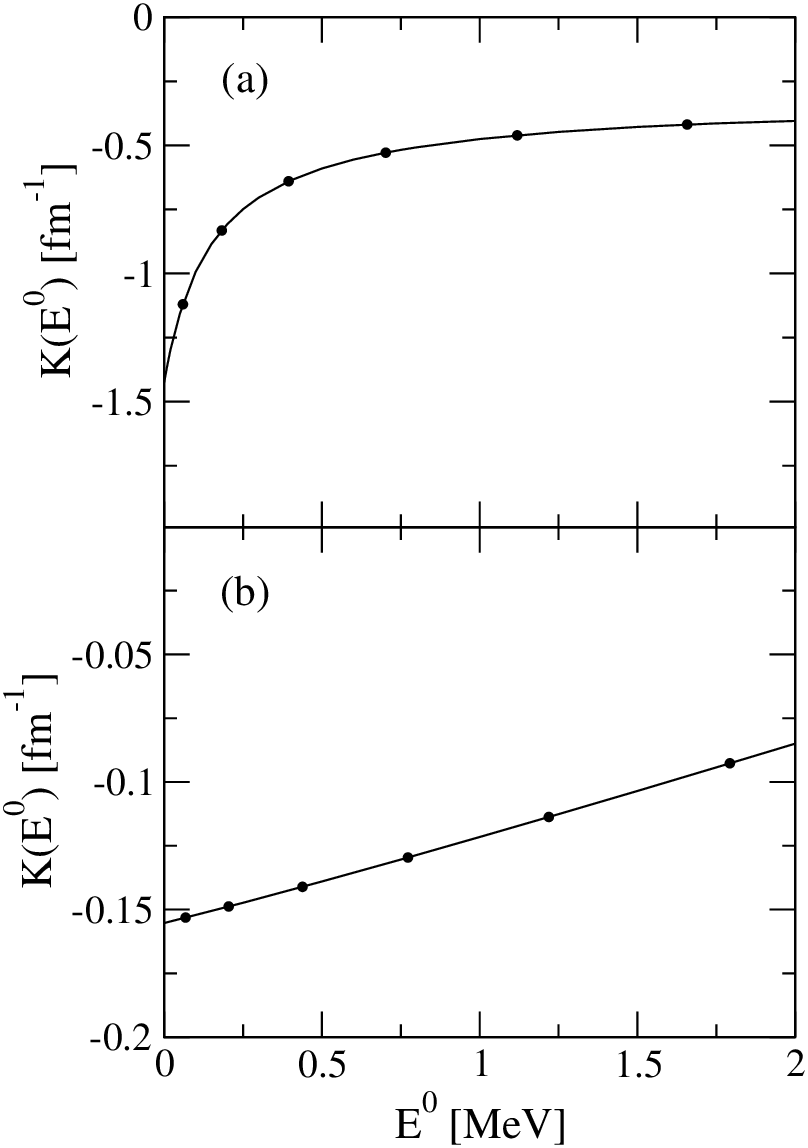,width=12cm}
\caption{The effective range function for $J=1/2^+$ (a) and $J=3/2^+$
(b) in the $n-d$ case. The solid points are obtained from the
second order estimates of $[\tan\delta_n]^{2^{nd}}$ given in Table~\protect\ref{tab:tab3}
for $^3$H, at the corresponding energies.}
\label{fig:fig4}
\end{figure}

\begin{figure}
\epsfig{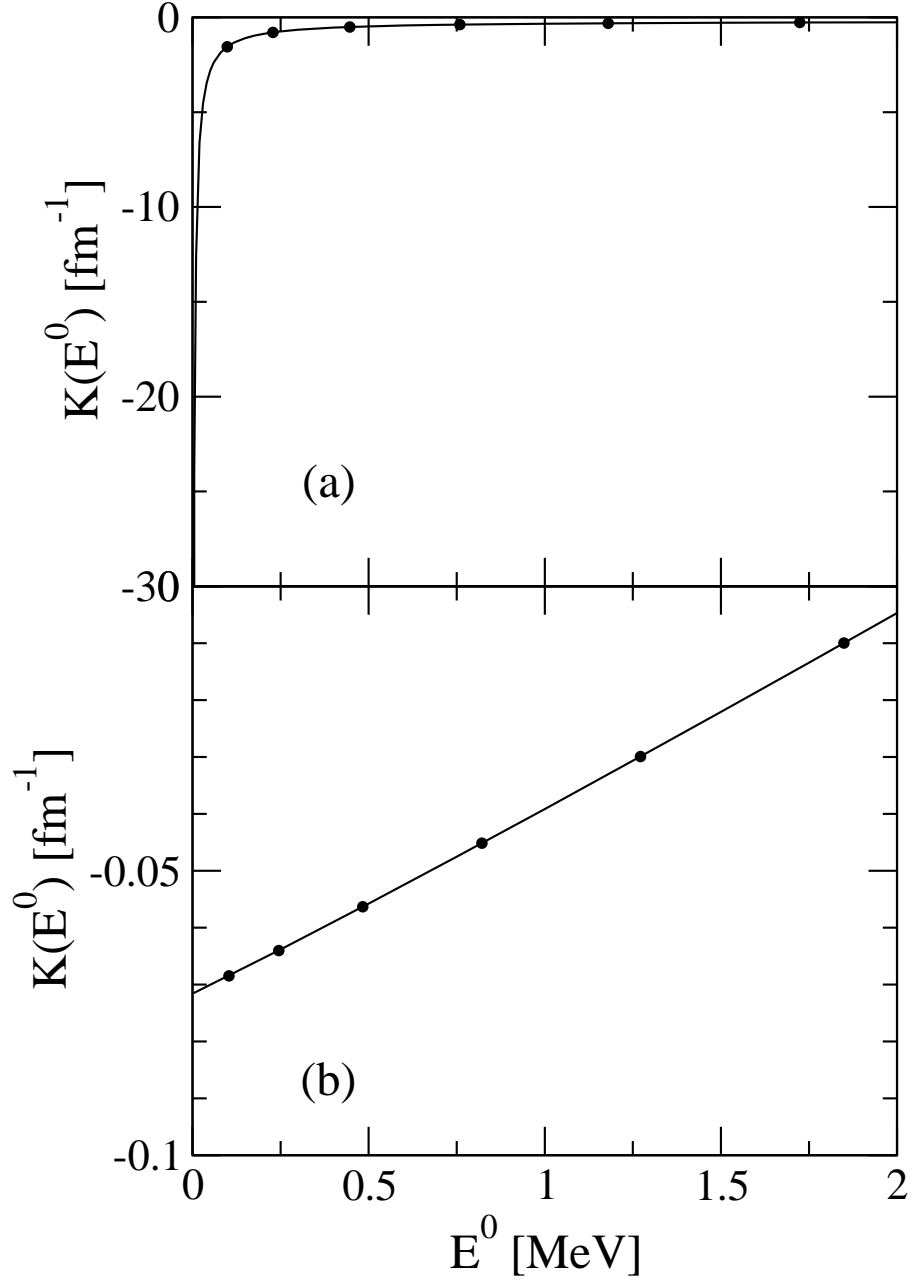}
\caption{The effective range function for $J=1/2^+$ (a) and $J=3/2^+$
(b) in the $p-d$ case. The solid points are obtained from the
second order estimates of $[\tan\delta_n]^{2^{nd}}$ given in Table\protect\ref{tab:tab3}
for $^3$He, at the corresponding energies.}
\label{fig:fig5}
\end{figure}

In Table~\ref{tab:tab2}, the $^3$H and $^3$He bound states and the doublet and quartet
n-d and p-d scattering lengths, corresponding to the MT I-III potential,
 are given in terms of the number $M$ of 
Laguerre polynomials used in the expansion of the hyperradial functions. 
The calculations have been done using $K=16$ which corresponds to 18 (9) hyperradial
functions in the case of $J=1/2^+$ ($J=3/2^+$). 
With $M=24$, an accuracy better than $1$ keV is obtained for
the bound state energies and of the order of $0.001$ fm for the scattering lengths. 
In Figs.~\ref{fig:fig4} and~\ref{fig:fig5} 
the $J=1/2^+,3/2^+$,  $l=0$, phase shifts $\delta$ are given as a function of the 
energy in the form of the effective range functions for $n-d$ and $p-d$, respectively. 
Following Ref.~\cite{chen89},
for $n-d$ scattering this function is defined as ($E^0=E-E_d$)
\begin{equation}
K(E^0)=k\cot\delta
\end{equation}
whereas for $p-d$ scattering it is defined as
\begin{equation}
K(E^0)=C_0^2(\eta)k\cot\delta+2k\eta h(\eta) \;\; ,
\end{equation}
where $\eta$ is the Coulomb parameter, $C_0^2=2\pi\eta/({\rm e}^{2\pi\eta}-1)$
and $h(\eta)=-{\rm ln}(\eta)+{\rm Re}\psi(1+i\eta)$ ($\psi$ is the digamma
function). The solid line in the figures represents these two functions obtained
solving Eqs.(\ref{linears},\ref{relintp}) for several values of the center of mass
energy $E^0$ in the interval $[0,|E_d|]$. The solid points
in the figures are the results obtained from the integral relations using
bound state wave functions as explained below.

The capability of the PHH to produce a very accurate description of bound
and scattering states can be used to study different applications
of the integral relations.
The lowest eigenvalue after the diagonalization procedure of Eq.~\refeq{geneig}
corresponds to the three-nucleon bound state energy of $^3$H ($T_z=-1/2$) or
$^3$He ($T_z=1/2$). However more negative eigenvalues could appear.
For example, in the case of $K=16$, $M=24$ and $\beta=1$ fm$^{-1}$,
six negative eigenvalues $E_n$ appear satisfying $|E_n|< |E_d|$.
They are given in Table~\ref{tab:tab3} transformed to the positive
energies $E_n^0=E_n-E_d$.
The corresponding eigenvectors $\Psi_n$ approximately describe a scattering process 
at the center of mass energy $E_n^0$, though asymptotically they
go to zero. In the following we use the index $n$ to
label these approximate scattering states and reserve the continuous index $k$ to label
those scattering states having the correct asymptotic behavior, as given by
Eq.~\refeq{asym1}. We now consider the diagonalization of
the Hamiltonian calculated using the PHH basis with the aforementioned values
of $K$, $M$, and $\beta$ but for the $J=3/2^+$ state.
The $J=3/2^+$ state does not have any bound state, however six negative
eigenvalues appear, all of them satisfying $|E_n|< |E_d|$. The positive 
energies $E_n^0$ are also given in Table~\ref{tab:tab3}. As in the previous
case, the corresponding eigenvectors approximately describe the $N-d$ scattering states, 
though asymptotically they go to zero. It should be observed that changing the values
of $K$, $M$ and $\beta$, the number of these states and the corresponding energies at 
which the eigenvalues appear change. They do not present the stability that a true
bound state shows. If we call $E_0^D$ the bound state energy calculated using
a basis of dimension $D$, the variational
principle establishes that $E_0^D\ge E_{3B}$, with $E_{3B}$ the energy
corresponding to $D\rightarrow\infty$. When the value of $D$ is sufficiently high,
a further increase of the dimension of the basis will not give an appreciable
improvement in $E_0^D$, showing a pattern of convergence of the type given
in Table~\ref{tab:tab2}. On the other hand, the eigenvalues $E_n$ are embedded
in the continuum spectrum of $H$ which starts at $E_d$. Accordingly,
increasing $D$ these eigenvalues tend to $E_d$ and the number of them also
increases. Similarly to what it has been done in the two-body case, we can
consider these states to be approximate solutions of $(H-E_n)\Psi_n=0$ in the interaction 
region and use them as inputs in the integral relation to compute 
second order estimates of the phase-shifts.
Defining
$ k^2_n=(4/3)E_n^0/(\hbar^2/m)$, with $\Psi_n$ the corresponding 
eigenvector and $|F_n>,|\tilde G_n>$ the asymptotic functions of Eq.~\refeq{asym1}
calculated at $k_n$, the second order estimate for the phase
shift at each energy is obtained as
\begin{equation}
\begin{aligned}
-&\frac{2m}{\hbar^2} <\Psi_n|H-E_n|F_n>=B_n  \cr
&\frac{2m}{\hbar^2} <\Psi_n|H-E_n|\tilde G_n>=A_n \cr
& [\tan\delta_n]^{2^{nd}} = \frac{B_n}{A_n}.
\end{aligned}
\label{rel5}
\end{equation}

The second order estimates of the phase-shifts for the six cases given in 
Table~\ref{tab:tab3} are shown in Figs.~\ref{fig:fig4} and \ref{fig:fig5} 
as solid points in the effective
range functions. We can observe an extremely good agreement with the scattering
calculations. This
method allows for the calculation of phase-shifts using bound state
type functions, even in the case of charged particles.
These results can be compared to the analysis of Ref.~\cite{chen89},
in which $N-d$ phase-shifts were obtained solving the Faddeev
equations in configuration space. For the $n-d$ case, the results presented
here and those from Ref.~\cite{chen89} are in complete agreement. In the
$p-d$ case the results of Ref.~\cite{chen89} were obtained considering
the Coulomb potential in $s$-wave, without including the correction obtained
using the KVP as was done in Ref.~\cite{friar83}. In fact, that paper reports
the doublet and quartet $p-d$ scattering lengths considering
the Coulomb potential in $s$-wave (the given values are
$0.16$ fm and $13.75$ fm, respectively).
After the correction introduced by using the KVP and considering the complete Coulomb 
potential, the results from Ref.~\cite{friar83} are $0.003$ fm and $13.95$ fm, respectively. 
They are in close agreement with the results obtained here and given
in Table~\ref{tab:tab2}. It is worth noting that the use of the integral relations
permits a correct computation of the $p-d$ phase-shifts in the energy range 
$[0,E_d]$,
after a diagonalization procedure of the Hamiltonian using square integrable basis
functions.

\begin{figure}
\vspace{2cm}
\epsfig{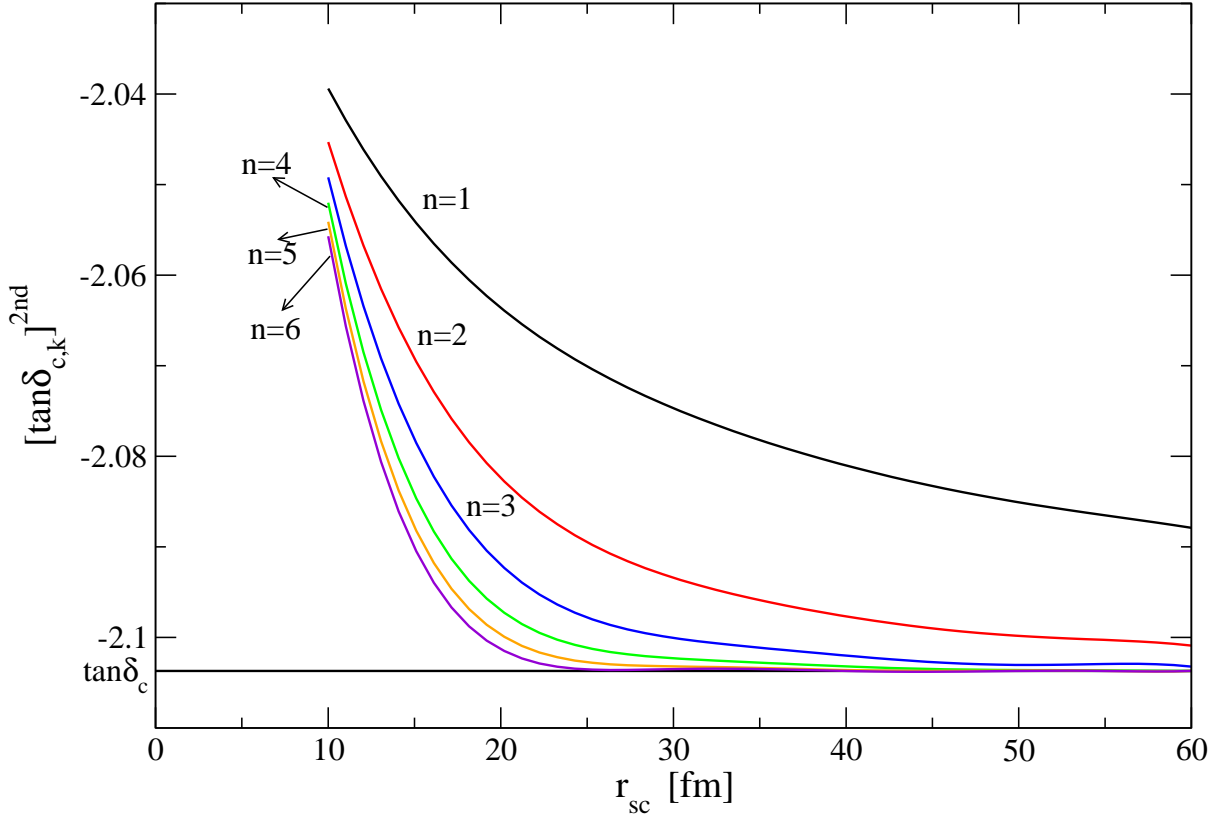}
\caption{(Color on line)
The three-nucleon second order estimate $[\tan\delta_{c,k}]^{2^{nd}}$
as a function of $r_{sc}$ for different values of $n$, at $E=0.2$ MeV.
As a reference
the exact value for $\tan\delta_c$ is given as a straight line. }
\label{fig:fig6}
\end{figure}

In the last example of this Section, we explore the possibility of extracting 
$p-d$ phase-shifts from a calculation in which the Coulomb potential has 
been screened at a certain distance, as we have already done for the two body case.
In the three nucleon system, we define the screened Coulomb potential as
\begin{equation}
 V_{sc}(i,j)= \left[{\rm e}^{-(r_{ij}/r_{sc})^n}\right]
\frac{e^2}{r_{ij}}(t^i_z+1/2)(t^j_z+1/2)
\label{coultrunc}
\end{equation}
with $r_{ij}$ the interparticle distance between nucleons $(i,j)$. Using
the PHH method, we solve
a $p-d$ scattering problem using the screened potential and, therefore, the
asymptotic behavior is of the form of Eq.~\refeq{asym1}, with $F_0$
and $G_0$ the regular and irregular Bessel functions. For different values
of $n$ and $r_{sc}$ we calculate the scattering wave function $\Psi^{n,r_{sc}}_k$
and, using the integral relations, we determine
the Coulomb phase shift. Similarly to what we have done in the two-body
case, the integral relations are
\begin{equation}
\begin{aligned}
 B^{2^{nd}}_k& = & -\frac{m}{\hbar^2}<&\Psi^{n,r_{sc}}_k|H-E|F_k>   \cr
 A_k & = & \frac{m}{\hbar^2}<&\Psi^{n,r_{sc}}_k|H-E|{\tilde G}_k>  \cr
 [\tan\delta_{c,k}]^{2^{nd}} & = & & B^{2^{nd}}_k/A_k  \,\, .
\end{aligned}
\label{relintp2}
\end{equation}
In $H$ the unscreened Coulomb potential is included and the asymptotic functions
$F_k$ and $\tilde G_k$ are given in Eq.~\refeq{asym1} in terms of the Coulomb 
functions $F_0,G_0$. The results
are shown in Fig.~\ref{fig:fig6} for the case of $J=3/2^+$ at $E_{cm}=2$ MeV.
The second order estimates  [$\tan\delta_{c,k}]^{2^{nd}}$ 
are given for different values of $n$ as a function
of $r_{sc}$. For $n>4$ and $r_{sc}\approx30$ fm, the results are in complete
agreement with the value obtained solving the $p-d$ case without any screening
of the Coulomb potential, $\tan\delta_c=-2.1037$, which is shown as a straight line
in the figure. This method can be
compared to the method given in Ref.~\cite{deltuva1} in which the Coulomb
potential was screened using the same screening function as in
Eq.~\refeq{coultrunc} and the Coulomb phase shift was recovered
after a renormalization procedure. We can conclude that
the integral relations used in Eq.~\refeq{relintp2} produce 
the same effect as the renormalization procedure.

\section{Integral Relations within the Hyperspherical Adiabatic Method}

In order to study applications of the integral relations using the HA method
for a three-nucleon system,
we give a brief introduction to the method following Refs.~\cite{intrel,barletta09} 
(for more details see Ref.~\cite{nielsen}). In the HA method the three-body wave
function of Eq.~\refeq{psi3b} is expanded as
\be
\Psi = \sum_{\mu=1}^{\infty} w_\mu(\rho) \Phi_\mu(\rho,\Omega),
\label{adbasis}
\ee
where $\Phi_\mu(\rho,\Omega)$ is a HA basis element and 
$[\rho,\Omega]\equiv [\rho,\phi_i,{\hat X}_i,{\hat Y}_i]$ is the set of
coordinates consisting of the hyperradius and of the five
hyperspherical coordinates. The HA basis elements are the eigenfunctions
of the hyperangular part of the Hamiltonian at fixed values of $\rho$:
\be
\left[\frac{\hbar^2}{2 m \rho^2}G^2 + V(\rho,\Omega)\right] \Phi_\mu(\rho,\Omega)=
U_\mu(\rho)\Phi_\mu(\rho,\Omega) \;\; ,
\label{adeigen}
\ee
where $G^2$ is the grand-angular operator and $V(\rho,\Omega)=\sum_i V(X_i)$ is
the potential energy. 
The eigenvalues, $U_\mu(\rho)$, are the adiabatic potentials, that appear
in the coupled set of differential equations
\begin{eqnarray}
& \left[ -\frac{\hbar^2}{2 m}T_\rho +U_\mu(\rho)-\frac{\hbar^2}{2 m}Q_{\mu\mu}(\rho)-E
 \right] w_\mu(\rho)-  \nonumber \\
&\frac{\hbar^2}{2 m}{\displaystyle \sum_{\mu'\ne\mu}^{N_A}} 
 \left[ Q_{\mu\mu'}(\rho) +P_{\mu\mu'}(\rho)(\frac{5}{\rho}+2 \frac{d}{d\rho})
\right]w_{\mu'}(\rho)=0
\label{usys}
\end{eqnarray}
with $T_\rho=\partial^2/\partial\rho^2+(5/\rho)\partial/\partial\rho$,
$N_A$ the number of adiabatic channels included in the calculation, $E$ the 
three-body energy, and from which the hyperradial functions $w_\mu(\rho)$ are obtained.
The coupling terms are defined as: 
$P_{\mu\mu'}=<\Phi_\mu(\rho,\Omega)|\partial/\partial\rho|\Phi_{\mu'}(\rho,\Omega)>$ and
$Q_{\mu\mu'}=<\Phi_\mu(\rho,\Omega)|\partial^2/\partial\rho^2|\Phi_{\mu'}(\rho,\Omega)>$.
In Ref.~\cite{intrel} the solution of the system of Eq.~\refeq{usys} has been studied
for scattering states below the three-body breakup. In that study it emerged that
the use of the integral relations helped to obtain a pattern of convergence for the 
phase-shift, in terms of the adiabatic channels, similar to the pattern obtained when 
the HA expansion is used to describe the bound states. 
It was also shown that the convergence, without the application
of the integral relations, is extremely slow. This problem originates in the boundary
conditions imposed on the solution of the linear system. As $\rho\rightarrow\infty$
the scattering wave function behaves asymptotically as
\be
 \Psi_k \rightarrow 
 \phi_d(r)\left[ \frac{\sin{(k_\rho \rho)}}{\sqrt{k_\rho}\rho}
    + \tan\delta_\rho\frac{\cos{(k_\rho \rho)}}{\sqrt{k_\rho}\rho}\right]|ST\ket .
\label{eqas}
\ee
with $|ST>$ the total spin-isospin function and 
$k^2_\rho=E^0/(\hbar^2/2m)=\frac{3}{2}k^2$. In fact,
when $\rho\rightarrow\infty$, the distance $X_i$ is limited by
$\phi_d$ and $y_i=\frac{\sqrt{6}}{2}Y_i\rightarrow\frac{\sqrt{6}}{2}\rho$, then
the approximate relation
$k y_i \approx k_\rho \rho$ holds. However, the exact equivalence
between $k y_i$ and $k_\rho \rho$ is not matched for
any finite value of $\rho$ and, accordingly, the boundary condition of
Eq.\refeq{eqas} is equivalent to that of Eq.\refeq{asym1} only at
$\rho\approx\infty$ and $N_A\rightarrow \infty$.
As a consequence $\delta_\rho$ converges extremely slowly to
$\delta$ by increasing the number of adiabatic states. Therefore,
the application of the integral relations in the case of the HA method, 
as discussed in Ref.~\cite{intrel}, removes the limitation given by the slow
convergence allowing an accurate description of the scattering states. 

\begin{figure}
\epsfig{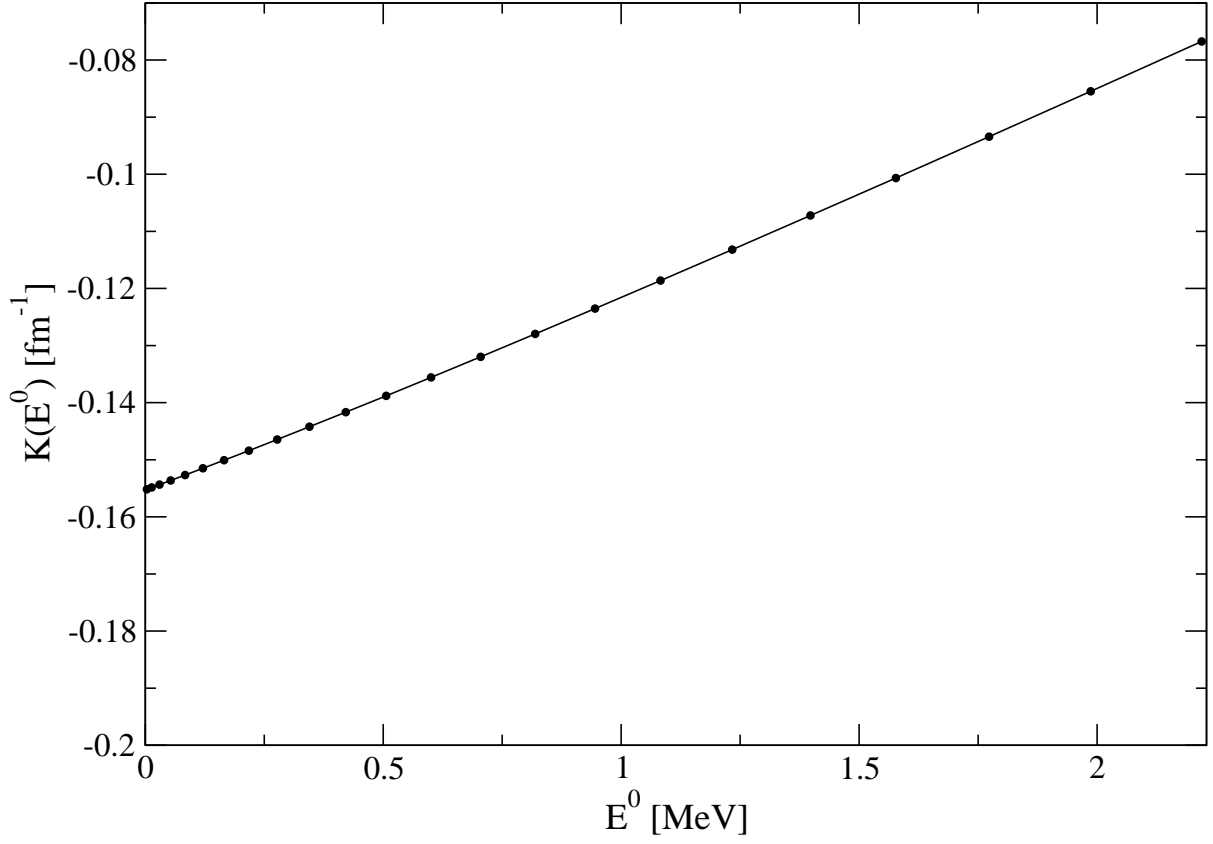}
\caption{The effective range function for $J=3/2^+$
(solid line) in the $n-d$ case. The solid points are obtained from the
second order estimates of $[\tan\delta_n]^{2^{nd}}$ using the HA expansion}
\label{fig:fig7}
\end{figure}

Motivated
by the results obtained in the previous section, we would like to analyze
the possibility of computing phase-shifts solving the system of Eqs.~\refeq{usys}
using bound state boundary conditions, namely imposing 
$w_\mu(\rho)\rightarrow 0$ as $\rho\rightarrow\infty$. To this end we expand the
hyperradial functions in the basis of Laguerre polynomial given in Eq.~\refeq{lag3}
and define the complete antisymmetric three-nucleon state $|\mu,m>$, with $\mu$
indicating a HA basis element and with $m$ a Laguerre basis element, respectively. 
In terms of this basis the three-body wave function results to be
\be
    \Psi_n=\sum_{\mu,m} A^{n}_{\mu,m} |\mu,m>
\ee
where $n$ indicates the different bound states.
As we did in the previous section, fixing the number of adiabatic channels $N_A$
and the number $M$ of Laguerre polynomials, we solve
the generalized eigenvalue problem for specific values of $J^\pi,T$ and identify
the negative eigenvalues $E_n$. They can represent true bound states 
($|E_n|>|E_d$) or they can indicate the possibility of approximate scattering
states ($|E_n|<|E_d$). As an example, for $J=3/2^+,T=1/2$, using
the MT-III potential with $N_A=40,M=60$ and with the non-linear
parameter fixed to the value $\beta=1$ fm$^{-1}$, a very dense spectrum 
of $23$ negative eigenvalues is obtained, all of them satisfying $|E_n|>|E_d$.
The corresponding
eigenvectors $\Psi_n$ are used to compute the second order estimates
of the phase-shifts at the specific energies using Eqs.~\refeq{rel5}. 
The results are shown
in Fig.~\ref{fig:fig7} as solid points on the effective range line. We observe a
perfect agreement between the $23$ points and the exact results given by
the straight line. In the solution of the same problem using the PHH basis, we
have used $M=24$ (see Table~\ref{tab:tab3}) and obtained six negative eigenvalues. 
With the HA expansion, using $M=60$, we obtain
a much denser spectrum covering the whole energy range $[0,E_d]$ below the
breakup into three particles.

\section{Conclusions}

The description of scattering states from the KVP has not been used in the literature
as much as the equivalent form for bound states, the Rayleight-Ritz variational
principle. A possible explanation for this is the different care that is required
at the moment of describing the asymptotic structure of
the system. For example, when a complete basis is used to describe an
$A$-body bound state, the main condition for the basis elements is that they
be square integrable. Elements having gaussian or exponential tails are
often used. It is well known that these bases do not describe
correctly the asymptotic structures as the distances between the particles
increase. However the error introduced by these configurations
in the computation of the binding energies is small. Conversely the extraction
of the asymptotic constants could be problematic if the number of basis states is
not sufficiently high (see for example Refs.~\cite{viviani05,kamimura}). The
situation drastically changes when the KVP is considered. The asymptotic structure
of the system has to be introduced in an exact form in the trial wave function
$\Psi_t$ otherwise the matrix element $<\Psi_t|H-E|\Psi_t>$ could well not be finite. 
In other words, $\Psi_t$ has to satisfy asymptotically that $(H-E)\Psi_t=0$. 
Formally $\Psi_t$ can be decomposed in an internal and in an asymptotic part 
as in Eq.~\refeq{psic}. Then, the internal part of $\Psi_t$
can be expanded over a square integrable basis. However, the necessity of taking
care explicitly of the (sometimes very complicated) asymptotic structures has
limited the application of the KVP. The reformulation of the KVP, given in
Eq.~\refeq{kohn}, to the form given in Eq.~\refeq{relint} changes this situation.
The KVP, given in terms of integral relations, does not necessitate the
explicit introduction of the correct asymptotic behavior in $\Psi_t$. The two
integrals involved, $<\Psi_t|H-E|F>$ and $<\Psi_t|H-E|\tilde G>$, go very fast
to zero since $F$ and $\tilde G$ are asymptotic solutions of the
Schr\"odinger equation. Therefore,
the only condition necessary to obtain accurate second order estimates through the
integral relations is that the trial wave function fulfill 
$(H-E)\Psi_t=0$ in the interaction region. This condition
can be achieved with a variety of methods.

In the present paper we have discussed two applications of the integral
relations:
the use of bound state like wave functions to describe scattering states
and, in the case of charged particles, the possibility of computing
phase-shifts using scattering wave functions with free asymptotic conditions,
obtained after screening the Coulomb interaction. Both problems are of
interest in the study of light nuclei. We started discussing the applications
to the $A=2$ system with a model (short-range) potential.
In this system the solution of the Schr\"odinger equation 
is possible and, therefore,
meaningful comparisons between the variational estimates
of the phase-shifts and the exact values can be performed. In the analysis
it was shown that after a diagonalization procedure of the two-nucleon
Hamiltonian those eigenvectors corresponding to eigenvalues embedded
in the continuum spectrum can be used as inputs in the integral
relations to determine the phase-shifts at those energies. We have
observed that increasing the number of basis states, the phase-shifts
converge to the exact values. In the second application we have
performed a scattering calculation adding to the short-range potential
a screened Coulomb potential.
Accordingly we have imposed free asymptotic conditions to the wave function.
It is well known that increasing the screening radius, the phase-shift
calculated with the screened potential will never match the phase-shift
obtained considering the full Coulomb potential. A renormalization
procedure is necessary (see Ref.~\cite{alt} and references therein).
It is very interesting to observe that the relation integrals as
given in Eq.~\refeq{rel4} produce the correct result. In fact, for 
suitable values of $r_{sc}$ and $n$,
the wave function calculated with the screened potential, 
$\Psi^{(n)}_{r_{sc}}$, is an approximate solution
of $(H-E)\Psi^{(n)}_{r_{sc}}=0$ in the
region in which the short-range potential is active, with $H$ containing
the bare Coulomb interaction. In fact, due to the short-range character 
of the integral relations, it is equivalent to use $\Psi^{(n)}_{r_{sc}}$ or the
wave function calculated with the Coulomb interaction in Eq.~\refeq{rel4}.

These examples have been discussed also in the three-nucleon system.
As a reference, we have used
the PHH method which gives a very accurate description
of the $A=3$ system and is well documented in the literature.
Firstly, we have calculated bound state wave functions 
using a semi-realistic interaction. For fixed values of $J^+$ and $T$ the
Hamiltonian has been diagonalized and attention has been given to
those eigenvalues satisfying $E_d<E<0$. This energy region corresponds to
$N-d$ elastic scattering and is located below the breakup into three
particles. The corresponding eigenvectors have been used to compute
the second order estimate of the tangent of the phase-shift.
In order to show the results in a visible way, we compute the
effective range function $K(E^0)$ using the PHH method, which gives an almost
exact result. Then, the second order estimates obtained from the bound
state like wave functions have been compared to $K(E^0)$.
We have observed that the variational estimates and the
exact results at the level of four digits coincide. This is practically the level
of accuracy reached by the PHH method, therefore we can conclude that
the results based on the integral relations can reach the same level
of accuracy as
other methods usually used to describe scattering states in $A=3$.
Moreover, a similar accuracy has been obtained
when the Coulomb interaction has been considered.
We consider this result to be of particular importance.
For example, the application of the Faddeev method for describing $p-d$ scattering
has been a subject of intense investigations and different techniques for including
the long range Coulomb interaction has been proposed (see the Introduction). 
From the results presented here it emerges that elastic $p-d$ scattering in the
low energy region  can be described 
using bound state like wave functions which can be easily computed.
Furthermore,
we have also analyzed the computation of $p-d$ phase-shifts from a 
calculation in which the Coulomb potential has been screened, as we 
have done in the two-nucleon system. Again, for suitable values of
$r_{sc}$ and $n$ we were able to reproduce the $p-d$ phase-shifts
using the integral relations. This result will be useful for a simple
extension of the Faddeev method, normally used to describe $n-d$
scattering, to determine $p-d$ phases without the normalization
procedure described in Ref.~\cite{deltuva1}. 

In the last Section we discuss an application of the integral relations
in connection with the HA method.
This method is often used to describe three-body bound states in nuclear,
atomic and molecular physics. It is very efficient, in particular 
when the interaction has a strong repulsion at short distances.
In Ref.~\cite{intrel} we have shown how to apply the HA method
to describe a $1+2$ collision solving the system of Eq.~\refeq{usys} 
with appropriate boundary conditions and then, using the integral relations,
the phase shift has been extracted. Here we have shown a different
application, the system of equations given in Eq.~\refeq{usys} has been 
solved using bound state
boundary conditions and then, using the integral relations,
the phase shift has been extracted. We have solved the same case 
previously considered
using the PHH expansion for the $J=3/2^+,T=1/2$ state. We have obtained
an extremely good description of the phase-shifts using HA bound state like
wave functions. This application will help to extend the applicability
of the HA to describe, for example, atom-dimer collisions at low energies.

Summarizing, we have demonstrated the usefulness of the KVP formulated in terms
of integral relations. We have shown the general validity of the formulation
with several applications to the $A=2,3$ systems.
In particular, we have in mind the possible use of this technique for 
describing scattering states using bound state methods in systems with $A\ge4$,
as for example the GFMC method or the
stochastic variational method.

\acknowledgments
This work was partly supported by DGI of MEC (Spain),
contract No.  FIS2008-01301. One of us (C.R.R.) thanks
support by a predoctoral I3P grant from CSIC and the European
Social Fund.

\newpage

\begin{table}[h]
\begin{tabular}{lcccc}
 M    &  10       &  20       &  30      &  40       \cr
\hline
$E_0$ &-0.395079 &-0.397740  &-0.397743 &-0.397743  \cr
\hline
$E_1$ & 0.536349 & 0.116356  & 0.048091 & 0.026008  \cr
$[\tan\delta_1]^{2^{nd}}$
      &-1.507280 &-0.622242  &-0.392005 &-0.286479  \cr
$\tan\delta_1$
      &-1.522377 &-0.621938  &-0.392021 &-0.286480  \cr
\hline
$E_2$ & 1.984580 & 0.449655  & 0.190019 & 0.103503  \cr
$[\tan\delta_2]^{2^{nd}}$
      &-5.919685 &-1.353736  &-0.812313 &-0.584389  \cr
$\tan\delta_2$
      &-5.703495 &-1.354691  &-0.812270 &-0.584388  \cr
\hline
$E_3$ & 4.512635 & 0.994433  & 0.423117 & 0.231645  \cr
$[\tan\delta_3]^{2^{nd}}$
      &13.998124 &-2.451174  &-1.302799 &-0.908128  \cr
$\tan\delta_3$
      &12.684474 &-2.448343  &-1.302887 &-0.908131  \cr
\hline
\end{tabular}
\caption{The two-nucleon bound state $E_0$ and the first three positive eigenvalues
$E_j$ $(j=1,3)$, as a function of the number of Laguerre polynomials $M$.
The second order estimates, $[\tan\delta_j]^{2^{nd}}$, obtained applying the integral
relations are given in each case and compared to the exact results, $\tan\delta_j$.}
\label{tab:tab1}
\end{table}

\begin{table}[h]
\begin{tabular}{lcc|cc|cc}
 $M$  &$B(^3{\rm H})$&$B(^3{\rm He})$&$^2a_{nd}$&$^4a_{nd}$&$^2a_{pd}$&$^4a_{pd}$ \cr  
\hline
 4    & -8.5117  & -7.8404 & 1.0207  & 6.4590 & 0.3987 &  14.088  \cr
 8    & -8.5351  & -7.8683 & 0.7251  & 6.4434 & 0.0363 &  13.978  \cr
 12   & -8.5357  & -7.8688 & 0.7031  & 6.4413 & 0.00636 & 13.967  \cr
 16   & -8.5357  & -7.8689 & 0.7019  & 6.4412 & 0.00472 & 13.966  \cr
 20   & -8.5357  & -7.8689 & 0.7018  & 6.4412 & 0.00461 & 13.965  \cr
 24   & -8.5357  & -7.8689 & 0.7018  & 6.4412 & 0.00458 & 13.965  \cr
 28   & -8.5357  & -7.8689 & 0.7018  & 6.4412 & 0.00456 & 13.965  \cr
 32   & -8.5357  & -7.8689 & 0.7018  & 6.4412 & 0.00454 & 13.965  \cr
\hline
\end{tabular}
\caption{Convergence of the $^3$H and $^3$He bound states (in MeV) and
the $n-d$ and $p-d$ doublet and quartet scattering lengths (in fm), 
using the PHH expansion, as a
function of the number of Laguerre polynomials $M$.}
\label{tab:tab2}
\end{table}

\begin{table}[h]
\begin{tabular}{cc|cc}
\hline
 \multicolumn{4}{c}{$^3$H}  \cr
\hline
 \multicolumn{2}{c|}{$J=1/2^+$}& \multicolumn{2}{c}{$J=3/2^+$} \cr
 $E_n^0$[MeV] &$[\tan\delta_n]^{2^{nd}}$&
 $E_n^0$[MeV] &$[\tan\delta_n]^{2^{nd}}$   \cr
\hline
   0.05934  & 0.03898 & 0.06789 & 0.30511 \cr
   0.18262  & 0.09204 & 0.20458 & 0.54508 \cr
   0.39445  & 0.17588 & 0.43850 & 0.84177 \cr
   0.70247  & 0.28429 & 0.77281 & 1.21661 \cr
   1.11898  & 0.41131 & 1.21923 & 1.74161 \cr
   1.65752  & 0.55093 & 1.79295 & 2.59081 \cr
\hline
 \multicolumn{4}{c}{$^3$He}  \cr
\hline
 \multicolumn{2}{c|}{$J=1/2^+$}& \multicolumn{2}{c}{$J=3/2^+$} \cr
 $E_n^0$[MeV] &$[\tan\delta_n]^{2^{nd}}$&
 $E_n^0$[MeV] &$[\tan\delta_n]^{2^{nd}}$   \cr
\hline
   0.09857  & 0.00753 & 0.10338 & 0.13807 \cr
   0.22822  & 0.03915 & 0.24470 & 0.33970 \cr
   0.44642  & 0.10965 & 0.48326 & 0.60099 \cr
   0.75903  & 0.21248 & 0.82144 & 0.91555 \cr
   1.18003  & 0.33871 & 1.27157 & 1.32466 \cr
   1.72398  & 0.48111 & 1.84927 & 1.92215 \cr
\hline
\end{tabular}
\caption{For the three-nucleon system, the six eigenvalues satisfying $E_d<E_n<0$
(given in the form $E_n^0=E_n-E_d$) in the
specific case of $K=16$, $M=24$ and $\beta=1$ fm$^{-1}$. The corresponding
second order estimate of $[\tan\delta_n]^{2^{nd}}$, obtained from the integral
relations, are also shown.}
\label{tab:tab3}
\end{table}

\newpage

\begin{thebibliography}{99}
\bibitem{nogga03} A. Nogga et al., Phys. Rev. {\bf C67}, 034004 (2003)
\bibitem{nogga02} A. Nogga, H. Kamada, W. Gl\"ockle, and B.R. Barrett,
 Phys. Rev. {\bf C65}, 054003 (2002)
\bibitem{viviani05} M. Viviani, A. Kievsky, and S. Rosati,
 Phys. Rev. {\bf C71}, 024006 (2005)
\bibitem{gloeckle}W. Gl\"ockle, H. Wita\l a, D. H\"uber, H. Kamada, and J. Golak,
 Phys. Rep. {\bf 274}, 107 (1994)
\bibitem{kievsky01}A. Kievsky, M. Viviani and S. Rosati,
        Phys. Rev. {\bf C64}, 024002 (2001)
\bibitem{fisher06}B.M. Fisher et al., Phys. Rev. {\bf C74}, 034001 (2006)
\bibitem{arnas07} A. Deltuva and A.C. Fonseca, Phys. Rev. {\bf C75},
       014005 (2007)
\bibitem{benchmark1}A. Kievsky et al., Phys. Rev. {\bf C58}, 3085 (1998)
\bibitem{benchmark2}R. Lazauskas et al., Phys. Rev. {\bf C71}, 034004 (2005)
\bibitem{gfmc}S.C. Pieper, K. Varga, and R.B. Wiringa,
 Phys. Rev. {\bf C66}, 044310 (2002)
\bibitem{ncsm}
P. Navr\'atil, V.G. Gueorguiev, J.P. Vary, W.E. Ormand, and A. Nogga, 
 Phys. Rev. Lett. {\bf 99}, 042501 (2007)
\bibitem{gfmc5} K.M. Nollett, S.C. Pieper, R.B. Wiringa, J. Carlson, and G.M. Hale,
 Phys. Rev. Lett. {\bf 99}, 022502 (2007)
\bibitem{ncsm5}S. Quaglioni and P. Navr\'atil,
Phys. Rev. {\bf C79}, 044606 (2009)
\bibitem{harris}F.E. Harris, Phys. Rev. Lett., {\bf 19}, 173 (1967)
\bibitem{suzuki}Y. Suzuki, W. Horiuchi, and K. Arai, Nucl. Phys. {\bf A823}, 1 (2009)
\bibitem{friar83}J.L. Friar, B.F. Gibson, and G.L. Payne,
Phys. Rev. {\bf C28}, 983 (1983)
\bibitem{friar84}J.L. Friar, B.F. Gibson, G.L. Payne, and C.R. Chen,
Phys. Rev. {\bf C30}, 1121 (1984)
\bibitem{phh}A. Kievsky, M. Viviani, and S. Rosati,
Nucl. Phys. {\bf A577}, 511 (1994)
\bibitem{kievsky01b}A. Kievsky, J.L. Friar, G.L. Payne, S. Rosati, and M. Viviani,
        Phys. Rev. {\bf C63}, 064004 (2001)
\bibitem{deltuva1} A. Deltuva, A.C. Fonseca, and P.U. Sauer,
        Phys. Rev. {\bf C71}, 054005 (2005)
\bibitem{alt} E.O. Alt, W. Sandhas, and H. Ziegelmann, Phys. Rev. {\bf C17}, 1981 (1978);
E.O. Alt and W. Sandhas, {\sl ibid} {\bf C21}, 1733 (1980)
\bibitem{intrel} P. Barletta, C. Romero-Redondo, A. Kievsky, M. Viviani, and E. Garrido,
Phys. Rev. Lett. {\bf 103}, 090402 (2009)
\bibitem{barletta09} P. Barletta and A. Kievsky, Few-Body Syst. {\bf 45}, 25 (2009)
\bibitem{mtiii} R.A Malfliet and J.A. Tjon, Nucl. Phys. {\bf A127}, 161 (1969)
\bibitem{phh1}A. Kievsky, M. Viviani and S. Rosati,
Nucl. Phys. {\bf A551}, 241 (1993)
\bibitem{kiev97}A. Kievsky, Nucl. Phys. {\bf A624}, 125 (1997)
\bibitem{chen89}C.R. Chen, G.L. Payne, J.L. Friar, and B.F. Gibson,
        Phys. Rev. {\bf C39}, 1261 (1989)
\bibitem{nielsen}E. Nielsen, D.V. Fedorov, A.S. Jensen, and E. Garrido,
Phys. Rep. {\bf 347} (2001) 373
\bibitem{kamimura} H. Kameyama, M. Kamimura and Y. Fukushima,
        Phys. Rev. {\bf C40}, 974 (1989)
\end{thebibliography}
\end{document}